\documentclass[final]{elsarticle_ARXIV}

\usepackage{graphicx}
\usepackage{amssymb}
\usepackage{amsmath}
\usepackage{euscript} 
\usepackage{gensymb}

\usepackage{xr}
\makeatletter
\newcommand*{\addFileDependency}[1]{
  \typeout{(#1)}
  \@addtofilelist{#1}
  \IfFileExists{#1}{}{\typeout{No file #1.}}
}
\makeatother
 
\newcommand*{\myexternaldocument}[1]{%
    \externaldocument{#1}%
    \addFileDependency{#1.tex}%
    \addFileDependency{#1.aux}%
}

\myexternaldocument{supp_mat}

\usepackage{geometry}

\begin{document}

\begin{frontmatter}


\title{
Immiscible fluid displacement in porous media with spatially correlated particle sizes}

\author[1]{Oshri Borgman\fnref{6}}
\author[3]{Thomas Darwent}
\author[2]{Enrico Segre}
\author[3]{Lucas Goehring}
\author[1]{Ran Holtzman\fnref{7}\corref{5}} \ead{holtzman.ran@mail.huji.ac.il}

\address[1]{Department of Soil and Water Sciences, The Hebrew University of Jerusalem, Rehovot 7610001, Israel}
\address[2]{Physics Core Facilities, Weizmann Institute of Science, Rehovot 7610001, Israel}
\address[3]{School of Science and Technology, Nottingham Trent University, Nottingham, NG11 8NS, United Kingdom}

\cortext[5]{Corresponding author}

\fntext[6]{Current address: The Department of Environmental Hydrology and Microbiology, Ben-Gurion University of the Negev, Sede Boqer 8499000, Israel}

\fntext[7]{Current address: Institute of Environmental Assessment and Water Research (IDAEA), Spanish National Research Council (CSIC), Barcelona 08034, Spain}

\begin{abstract}
Immiscible fluid displacement in porous media is fundamental for many environmental processes, including infiltration of water in soils, groundwater remediation, enhanced recovery of hydrocarbons and CO\(_2\) geosequestration.
Microstructural heterogeneity, in particular of particle sizes, can significantly impact immiscible displacement. For instance, it may lead to unstable flow and preferential displacement patterns.
We present a systematic, quantitative pore-scale study of the impact of spatial correlations in particle sizes on the drainage of a partially-wetting fluid.
We perform pore-network simulations with varying flow rates and different degrees of spatial correlation, complemented with microfluidic experiments.
Simulated and experimental displacement patterns show that spatial correlation leads to more preferential invasion, with reduced trapping of the defending fluid, especially at low flow rates.
Numerically, we find that increasing the correlation length reduces the fluid-fluid interfacial area and the trapping of the defending fluid, and increases the invasion pattern asymmetry and selectivity.
Our experiments, conducted for low capillary numbers, support these findings.
Our results delineate the significant effect of spatial correlations on fluid displacement in porous media, of relevance to a wide range of natural and engineered processes.
\end{abstract}


\end{frontmatter}

 \newpage

\section{Introduction} \label{sec:Intro}

Fluid displacement plays a key role in many natural and engineered environmental applications \citep{Blunt2017}, for example the infiltration of water into soil \citep{Dagan1983}, groundwater contamination and soil remediation \citep{Johnson1993,McCray2000}, enhanced hydrocarbon recovery \citep{Alvarado2010} and CO$_2$ sequestration \citep{Herring2013}.
While these processes are typically observed and modeled over large spatial scales (meters and above), the physical behavior at the pore-scale is crucial to understanding and predicting emergent behavior and the selection of particular flow patterns, such as fingering~\citep{Toussaint2012,Picchi2018}.
Structural heterogeneity---an inherent feature of porous and fractured media such as soils, sediments and rocks \citep{Knackstedt2001,Trevisan2015a,Ye2015}---strongly impacts fluid displacement.
Specifically, it can lead to unstable or preferential flows, which affect processes such as water redistribution in soil~\citep{Schluter2012}, pressure-saturation relationships in granular media \citep{Murison2014}, evaporative drying \citep{Borgman2017}, solute transport~\citep{Amooie2017} and even the distribution of fresh and saline groundwater at the continental shelf~\citep{Michael2016}.
Recent works indicate the increasing interest in the role of the microstructure of a porous medium in fluid displacement and transport processes within it \citep{Alim2017,Borgman2017,Nissan2018}. For instance, it has been shown that the local correlations between pore sizes can have a greater effect on flow velocities than the pore size distribution itself~\citep{Alim2017}.

Fluid invasion patterns are determined by the competition between a number of processes and flow parameters, including gravity, the fluid viscosity and the wettability of the solid material, as well as pore sizes and internal topology \citep{Or2008,Toussaint2012,Holtzman2015}.
Here, in the absence of gravity and wettability effects (namely, for a strongly wetting defending fluid), the displacement patterns are controlled by two dimensionless groups:
the viscosity ratio $\mathcal{M} = \mu_i/\mu_d$, where $\mu_i$ and $\mu_d$ are the viscosities of the invading and defending fluid, respectively, and the capillary number $\mathrm{Ca}=\mu_dv/\sigma$, where $v$ and $\sigma$ are the characteristic liquid velocity and interfacial tension, respectively \citep{Blunt2017}.
In drainage, for $\mathcal{M} < 1$, displacement patterns vary from capillary fingering (CF) at low $\mathrm{Ca}$ to viscous fingering (VF) at high $\mathrm{Ca}$.
When $\mathcal{M} \ll 1$, viscous instabilities dominate and VF patterns are formed regardless of the flow rate~\citep{Lenormand1988}.
Morphologically, CF patterns approach a fractal shape, and are characterized by many trapped defending fluid clusters, whereas VF patterns exhibit thin, branching fingers, with fewer interconnections and, hence, reduced trapping ~\citep{Meheust2002,Holtzman2016}.
Reducing the wettability of the defending fluid leads to cooperative pore filling, which results in a more compact pattern with a smoother fluid-fluid interface at low \(\mathrm{Ca}\) and \(\mathcal{M} < 1\) \citep{Trojer2015,Holtzman2015,Holtzman2016}.

The microstructure of a porous medium, characterized by both the distribution and the spatial arrangement of pores of various sizes, also has a substantial impact on fluid displacement patterns~\citep{Ferrari2015,Tsuji2016,Holtzman2016,Fantinel2017a}.
For example, the transition between VF and CF depends on the statistical distribution of pore sizes.
A broader pore-size distribution increases local differences in capillary thresholds, thus overcoming the viscous pressure drop for longer ranges, and maintaining CF patterns for higher flow rates \citep{Holtzman2010,Ferrari2015}.
High disorder also leads to a larger interfacial area between the fluids during both drainage and weak imbibition, with more trapping of the defending fluid~\citep{Holtzman2016}.  In other words, disorder works to stabilize the invasion front. 
Even for unstable viscosity ratio \(\mathcal{M} < 1\), changing the pore geometry, for example by introducing a gradient in pore size, can stabilize the displacement front~\citep{Al-Housseiny2012,Rabbani2018}.

An important feature of the microstructure of many types of porous media such as soils and rocks is the existence of spatial correlations in pore sizes, such as when pores of similar size are clustered together, creating distinct regions with different hydraulic properties. 
The impact of these correlations on drainage patterns has been studied mainly in the context of quasi-static displacement, where it was shown to cause a smoother displacement front~\citep{Paterson1996,Knackstedt2001}.
In addition, long-range correlations lower the percolation threshold~\citep{Renault1991} and the saturation of the invading phase at breakthrough~\citep{Paterson1996}, decrease the residual saturation thus altering the pressure-saturation relationship \citep{Rajaram1997,Knackstedt2001}, and increase the relative permeability of both the wetting and non-wetting phases \citep{Mani1999}. Furthermore, we recently demonstrated the impact of spatial correlations on drying---a  fluid displacement process driven by the evaporation of the defending liquid---showing that pore-size correlations lead to the preferential invasion of connected regions of large pores, and a prolonged high-drying-rate period \citep{Borgman2017}, with more pronounced and intermittent pressure fluctuations \cite{Biswas2018}. While these studies shed important light on correlation effects at the quasi-static limit, their interplay with rates, namely dynamic (or viscous) effects, have not yet been considered.

Here, we systematically study the impact of spatial correlations in pore sizes on fluid displacement, using pore-scale simulations combined with state-of-the-art microfluidic experiments.
We show that increasing the correlation length (the characteristic size of patches of pores with similar sizes) leads to more preferential invasion, reduced trapping, a smoother fluid-fluid interfacial area and reduced sweep efficiency. We also show that these effects diminish at high flow rates, when viscous forces become dominant. 

\section{Methods} \label{sec:Methods:Sims}
\begin{figure}[ht!]
\centering
\noindent\includegraphics[width=.95\columnwidth]{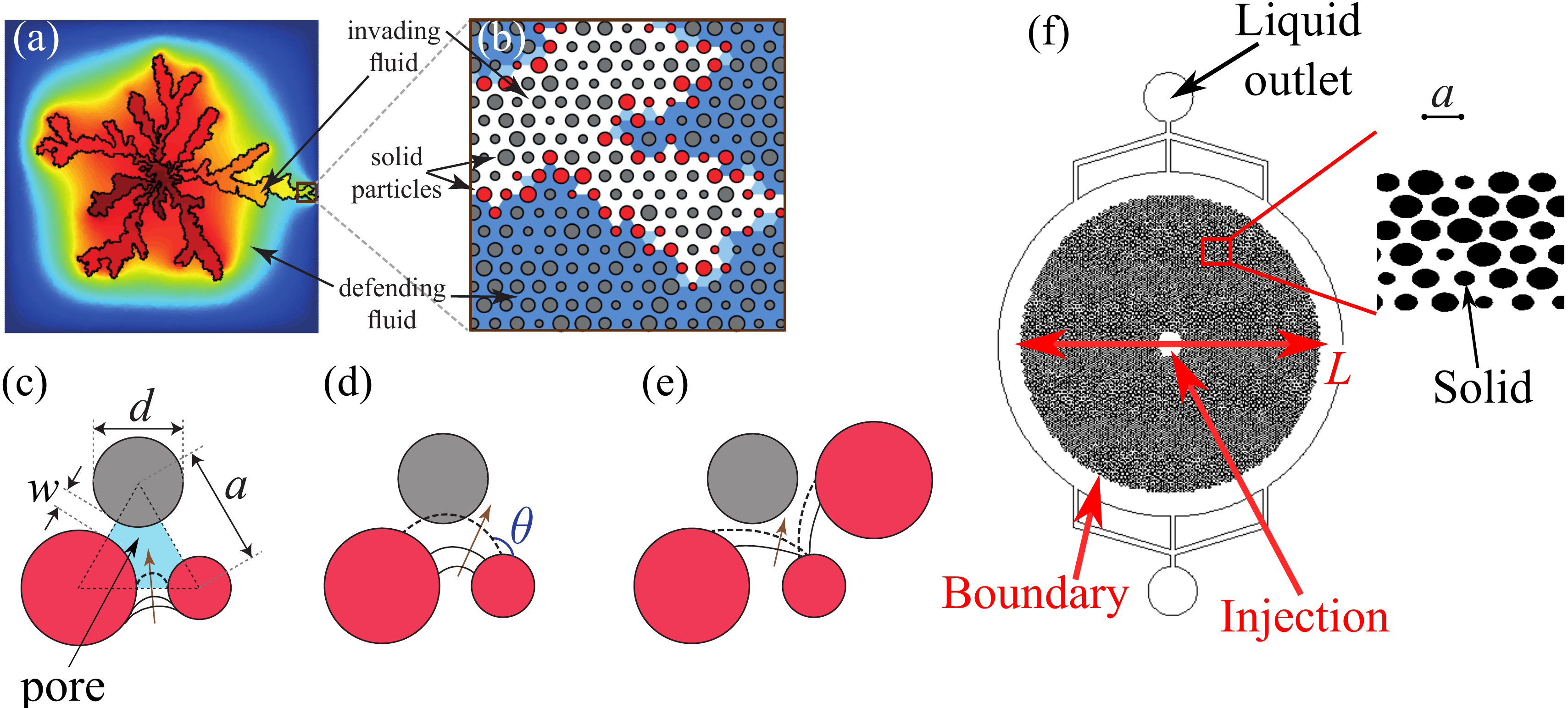} 
\caption{Our porous medium is made of variably-sized cylindrical pillars placed on a triangular lattice. 
(a) Numerically, we track the fluid-fluid interface (black line) and fluid pressures (increasing from blue to red) during immiscible fluid invasion.
(b) A close-up view shows the lattice (of spacing $a$) of particles (of variable diameter $d$) and the pores interconnected by throats (of width $w$).
The fluid-fluid interface is represented as a sequence of circular menisci, touching particles at contact angle $\theta$, with radii of curvature $R$ set by the local capillary pressure.
Menisci can be destabilized by:
(c) {burst}, (d) {touch}, or (e) overlap.
In each sketch the brown arrows indicate the direction of advancement of the meniscus, and destabilized menisci are represented by dashed arcs (Reproduced from~\citep{Holtzman2015}; Copyright 2015 by the American Physical Society). 
(f) Experimental microfluidic cell design, showing a central air injection site, and a peripheral boundary zone connected to the liquid outlet ports.
A close-up view shows the variation in the solid particles (pillars).
\label{fig:Methods}
}
\end{figure}

The complexity of flows in natural porous materials, such as soils and rocks, makes understanding their underlying physics a challenging task.
Thus, we consider---experimentally and numerically---an analog porous medium with a simplified geometry:
an array of solid, cylindrical pillars on a triangular lattice, where heterogeneity is provided by variations in the pillar radii (Fig.~\ref{fig:Methods}).
Specifically, we investigate the radial invasion patterns of fluids flowing into circular patches of such pillars, in order to simplify the pattern characterization.
However, we note that our methodologies could allow for a wide variety of designs, for instance other ordered lattices \cite{Fantinel2017a} or random pillar arrangements.

Our simulations are based on the model of~\citet{Holtzman2015}, and are compared with microfluidic experiments of similar pore geometry.
The model provides a mechanistic description of partially-wetting fluid-fluid displacement, and represents the basic interplay between capillary and viscous forces, invasion dynamics and wettability.
It is also computationally efficient, allowing us to rapidly conduct many realizations in a large domain size.
The experiments provide pore-scale observations of exceptional resolution, which enable us to verify our modeling results and provide better insight on the pore-scale physics.

\subsection{Pore-scale model} \label{sec:Methods:Sims:Model}
Here we describe our numerical model, as sketched in Fig. 1(a-e); for further details see \citet{Holtzman2015}. The model is a hybridization of two complementary pore-scale modeling approaches: pore-based and grain-based.
Pore-network models resolve pore pressures and interpore fluxes from pore topology and geometry~\citep{Joekar-Niasar2012}, while grain-based models incorporate the different pore filling mechanisms that arise due to wettability effects, by linking the meniscus geometry to the local capillary pressure, grain size and contact angle~\citep{Cieplak1990}. 
An advantage of our model is its ability to capture the impact of flow dynamics (in particular, fluid viscosity effects and meniscus readjustments, as in~\citet{Furuberg1996}), along with the impact of partial wettability on pore-scale displacement mechanisms \citep{Holtzman2015}.
The model does not include the effects of additional mechanisms such as droplet fragmentation, snapoff, or film flow.

The basic status of each pore is determined by its fluid content, $\varPhi$, where \(\varPhi = 0\) or 1 for a pore which is completely filled by the  defending or invading fluid, respectively.  The invasion front is defined by the interface separating fully invaded pores, where $\varPhi =1$, from accessible non-invaded ($\varPhi =0$) or partially invaded ($0<\varPhi <1$) pores.
A pore is considered accessible if it is topologically connected through the defending fluid to the outer boundary;
as a result, the volume of the trapped clusters of the defending fluid is fixed, and they cannot be invaded or refilled in our simulations.
Note also that with this, we do not consider connectivity of the wetting fluid through films.

Along the invasion front the fluid-fluid interface is approximated by a sequence of menisci, shaped as circular arcs.
Each arc intersects a pair of particles at the prescribed contact angle $\theta$ and has a curvature
\begin{equation}
\kappa = 1/R=  \Delta p /\sigma,     
\end{equation}
where $R$ is the radius of curvature of the meniscus, and $\Delta p$ is the capillary pressure (the pressure jump across the meniscus), computed from the Young-Laplace law. 
We consider a cell filled with pillars whose height, $h$, is large compared to the throat aperture $w$, such that $h\gg w$. 
This allows us to use a two-dimensional model, where $R$ can be estimated from the in-plane curvature alone.
The angle $\theta$ is measured through the defending fluid (i.e. $\theta< 90^{\circ}$ for drainage), and represents an \textit{effective} advancing contact angle, including any dynamic effects~\citep{Meakin2009}.
Knowledge of $R$ and $\theta$ allows us to analytically resolve the geometry, and hence stability, of each meniscus.  Specifically, the menisci are tested for three types of capillary instabilities~\citep{Cieplak1990}, as sketched in Fig.~\ref{fig:Methods}(c--e): (1) a
Haines jump or \emph{burst}, when the curvature $\kappa$ exceeds a threshold set by the local geometry;
(2) \emph{touch}, when a meniscus makes contact with a downstream particle; and
(3) the \emph{overlap} of adjacent menisci, intersecting and hence destabilizing each other. 

An unstable meniscus is allowed to advance into its downstream pore, at a rate which depends on the viscous dissipation due to its surrounding pore structure (the hydraulic resistance of the constriction).
Here, fluid pressures are evaluated by simultaneously resolving the flow between all pores containing the same phase, 
and through any throat with an advancing meniscus. 
We also solve for the flow of both the invading and the defending fluids, at the same time.
Mathematically this is done by assuming incompressible flow and enforcing in each pore the conservation of mass,  
\begin{equation}
\Sigma_j q_{ij}=0,
\end{equation}
by summing the fluxes $q_{ij}$ into pore $i$ from all connected pores $j$.
This provides a system of linear equations that is solved explicitly at each time step.
The volumetric flow rate between neighboring pores is evaluated by
\begin{equation}
q_{ij} = C_{ij} \nabla p_{ij},
\end{equation}
where $C_{ij}\sim w_{ij}^4/\mu_{eff}$ is the conductance of the connecting throat.
The effective viscosity $\mu_{eff}=\left(\mu_i-\mu_d\right)\varPhi+\mu_d$, where $\mu_i$ and $\mu_d$ are the viscosities of the invading and defending phases, respectively.
This provides the flow rate $q_{ij}$ between pores containing the same fluid as well as the meniscus advancement rate in partially-filled pores, in which the meniscus is unstable.
Note that in pores with stable menisci, we ignore the small volume changes associated with meniscus curvature variations (due to changes in capillary pressure).
The local pressure gradient, $\nabla p_{ij}=\left(p_j-p_i\right)/\Delta x_{ij}$, is evaluated from the pressure difference between the two neighboring pores.
If these pores contain different fluids, the pressure difference is simply the capillary pressure.
In this calculation we rely on the assumption that most of the resistance to flow occurs in the narrow constriction between two pores (i.e., the throat), therefore $\Delta x_{ij}=w_{ij}$.

To track the progression of the fluid invasion, at each time step we:
(1) locate the position of the invasion front from the filling status $\varPhi$ and define the connected networks of pores within both fluids;
(2) evaluate the pressure $p$ in each pore, and calculate the flow rate $q$ for each throat;
(3) check for new meniscus instabilities and update the flow network accordingly;
and (4) update the filling status of each invaded pore by $\varPhi \left(t+\Delta t\right) = \varPhi \left(t\right) + q^{inv} \left(t\right) \Delta t /V$, where $q^{inv}$ is the inflow of invading fluid, and $V$ is the pore volume.

At the source of the invading fluid we enforce a constant injection rate by setting the hydraulic resistance of an `inlet' region---a disk with a diameter covering about 12 pores, surrounding the central injection point---to be orders of magnitude larger than elsewhere, and by keeping fixed the pressure drop between the inlet and the ring of outermost (outlet) pores.
This ensures a practically constant pressure gradient that is maintained throughout the simulation, regardless of the front position, and a nearly constant flux of fluid into the sample.
Simulations are terminated by the breakthrough of the invading fluid, i.e. once any pore on the outer boundary is invaded.

The time-step $\Delta t$ is chosen so that only a small fraction of a pore (not more than 30\% of any invaded pore) may be filled by the invading fluid, in each step.
When a pore invasion ends (i.e. when $\varPhi=1$) the interface configuration is updated by replacing any unstable menisci with new ones that touch the particles upstream from the newly invaded pore.
The finite pore filling time in our model, while allowing pores which are partially-filled to re-empty if the direction of meniscus advancement is reversed, enables our model to capture dynamic (viscous) effects, overcoming a long-standing computational challenge~\citep{Meakin2009,Joekar-Niasar2012}. These effects include pressure screening~\citep{Lovoll2004} and interface readjustments: the non-local decrease in menisci curvature following a pore invasion, due to the redistribution of the defending fluid~\citep{Furuberg1996,Armstrong2013}. 

\subsection{Sample geometry and simulation parameters} 
\label{sec:Methods:Sims:Parameters}
Our sample geometry is a circular cell of diameter $L = 120a$, containing cylindrical pillars on a traingular lattice, where $a=45~ \mu \mathrm{m}$ is the lattice length (the distance between the centers of two adjacent pillars).  The pillars, as well as the cell, have a height of $h = 65 ~\mu \mathrm{m}$.
Their diameters $d$ have a mean size $\overline{d}=25~ \mu \mathrm{m}$ and standard deviation $\sigma_d=5~ \mu \mathrm{m}$, arranged in a spatially correlated pattern (see below).
The values of $d$ are limited to the range [$d\left(1-\lambda\right)$, $d\left(1+\lambda\right)$], where $\lambda=0.8$; this constraint prohibits blocked throats due to particle overlaps.
A pore is defined as the open volume between a set of three adjacent pillar centers, and a throat is defined as the constriction between two adjacent pillars. The throat apertures thus have a mean size of $\overline{w}=20~ \mu \mathrm{m}$, and the pore volumes are related to their surrounding pillar and throat sizes.

To introduce spatial correlations in pillar sizes, we generate a random rough surface $H(x,y)$ such that its Fourier transform is a Gaussian distribution of intensities, centered around zero, with random phases.
This is prepared by summing $10^4$ sinusoidal waves, whose amplitude, phase and orientation are selected from random uniform distributions, and whose wave numbers were drawn from a normal distribution.
The width of this distribution, in the Fourier domain, is inversely proportional to the correlation length $\zeta$ of the surface (in units of the lattice length).
For a review of the methods to generate such rough surfaces see~\citet{Persson2005}.
The diameter of each pillar is now defined such that $d_i = \overline{d}(1 + H_i)$, where $H_i$ is the height of the correlated surface at a specific pillar coordinate ($x_i,y_i$).
To obtain statistically-representative results from the simulations, averages and deviations for various metrics of the displacement patterns were computed from an ensemble of 10 realizations (namely, samples with the same statistical attributes but with different random seeds), for each $\zeta \in \{1, 2, 3.5, 5, 10\}$.

We vary the $\mathrm{Ca}$ in the simulations by varying the inlet pressure.
The average flow rate is calculated as $Q=V_{tot}/t_{tot}$, with $V_{tot}$ and $t_{tot}$ being the total displaced volume and time at breakthrough, respectively.
This then provides a characteristic velocity of $v = Q/A_{out}$, where $A_{out}$ is the cross-sectional area of the cell's outer perimeter, which is open to flow (i.e. the sum of cross-sectional area of pore throats on the perimeter). 
Other parameters used in the simulations are the interfacial tension 
$\sigma = 71.67\times 10^{-3} ~N/\mathrm{m}$, 
and the viscosities of the invading and defending fluids, 
$\mu_i = 1.8\times 10^{-5}~$Pa~s
and
$\mu_d = 1\times 10^{-3}~$Pa~s,
respectively.    
These values model the displacement of water by air.
Finally, the contact angle in the simulations was set to $\theta =$ 73\degree, to match the experimental conditions (see following section).

\subsection{Microfluidic displacement experiments} \label{sec:Methods:Exp}

Microfluidic micromodels are produced using soft lithography techniques as detailed further by \citet{Madou2002} and \citet{Fantinel2017a}.  We use a high-resolution chrome-quartz photomask and a negative photosresist (SU8 3025) to manufacture reusable templates.
Poly(dimethylsiloxane) (PDMS) is poured over these master templates, degassed under vacuum, and cured for 1 h at 75\degree C.
The PDMS covering a designed pattern is then cut and peeled off the master, and inlet/outlet holes punched.  
This patterned slab, and a PDMS-coated glass slide, are primed in an oxygen plasma and adhered to one another, forming a microfluidic chip of solid pillars separated by open channels with a thickness of 65$~\mu\mathrm{m}\pm 3 ~\mu \mathrm{m}$.  An example design is shown in Fig. \ref{fig:Methods}(f).

The experiments use a similar geometry to the numerical simulations, but are scaled slightly as a result of practical considerations.  
The lattice length in the experimental cells is $a=130~\mu \mathrm{m}$, the pillar mean diameter $\overline{d}=80~\mu \mathrm{m}$ and the pillar size disorder is given by $\lambda=0.5$, with spatial correlation lengths of $\zeta \in \{1,2, 3.5, 5, 10\}$.
These values allow for a minimum throat width of $10~\mu \mathrm{m}$, which ensures for the reliable fabrication of the prepared designs, while maintaining a high degree of size heterogeneity.
By observing the cells under a microscope, we estimate an uncertainty in pillar diameter of $\sim$1.0 $\mu$m (i.e. the pillar is within $\sim$1.25\% of the designed size). 

PTFE tubes (Adtech Polymer Engineering Ltd) are inserted into the inlet and outlets, and are sealed in place with a UV-curing adhesive (AA 3526, Loctite).
A water/glycerol mixture is then pumped through the central inlet to fill the cell.
Any trapped air bubbles disappear after a few minutes, leaving the cell saturated with liquid.
The cell is then mounted horizontally under a digital camera (Nikon D5100) with the glass slide facing upwards.
The entire apparatus is housed in a darkened box, with a low-angle LED strip surrounding the cell to highlight interfacial features.
A syringe pump then withdraws liquid out of the cell through the outlet ports on its perimeter, allowing air to invade via the central inlet (see Fig. \ref{fig:Methods}f).
Time-lapse images are taken every second until the liquid-air interface reaches breakthrough. 
The syringe pump withdrawal rate is fixed at $Q=1.30\times 10^{-10}$ m$^3$/s, and $\mathrm{Ca}$ is varied by changing the composition (and hence viscosity) of the water/glycerol mixture to give Ca values of 1.9$\times 10^{-6}$, 3.2$\times 10^{-6}$, 1.3$\times 10^{-5}$ and 3.7$\times 10^{-5}$.  Overall, 24 experiments were performed with various $\mathrm{Ca}$ and $\zeta$.
The receding contact angle between the water/glycerol mixtures and PDMS was measured as 73\degree $~\pm$ 8\degree\ using a drop shape analyzer (DSA-10, Kr\"{u}ss Scientific).

A summary of the image analysis process applied to the microfluidic experiments is provided in the Supplementary Material. 

\section{Results}
\label{sec:Results}

We find that at low to intermediate capillary numbers $\mathrm{Ca}$ (i.e. slow, low viscosity flows), increasing the correlation length of the disorder promotes preferential fluid displacement patterns.
Less of the defending fluid is flushed out before breakthrough, but also less volume is trapped behind the invasion front. Essentially, invasion is forced to occur through connected patches of large pores, following the path of least resistance.
At high $\mathrm{Ca}$, fluid displacement is instead controlled by viscous dissipation rather than capillary forces, and the impact of the spatial correlation on the invasion patterns is less apparent.

\begin{figure}[t!]
\centering
\noindent\includegraphics[width=1\columnwidth]{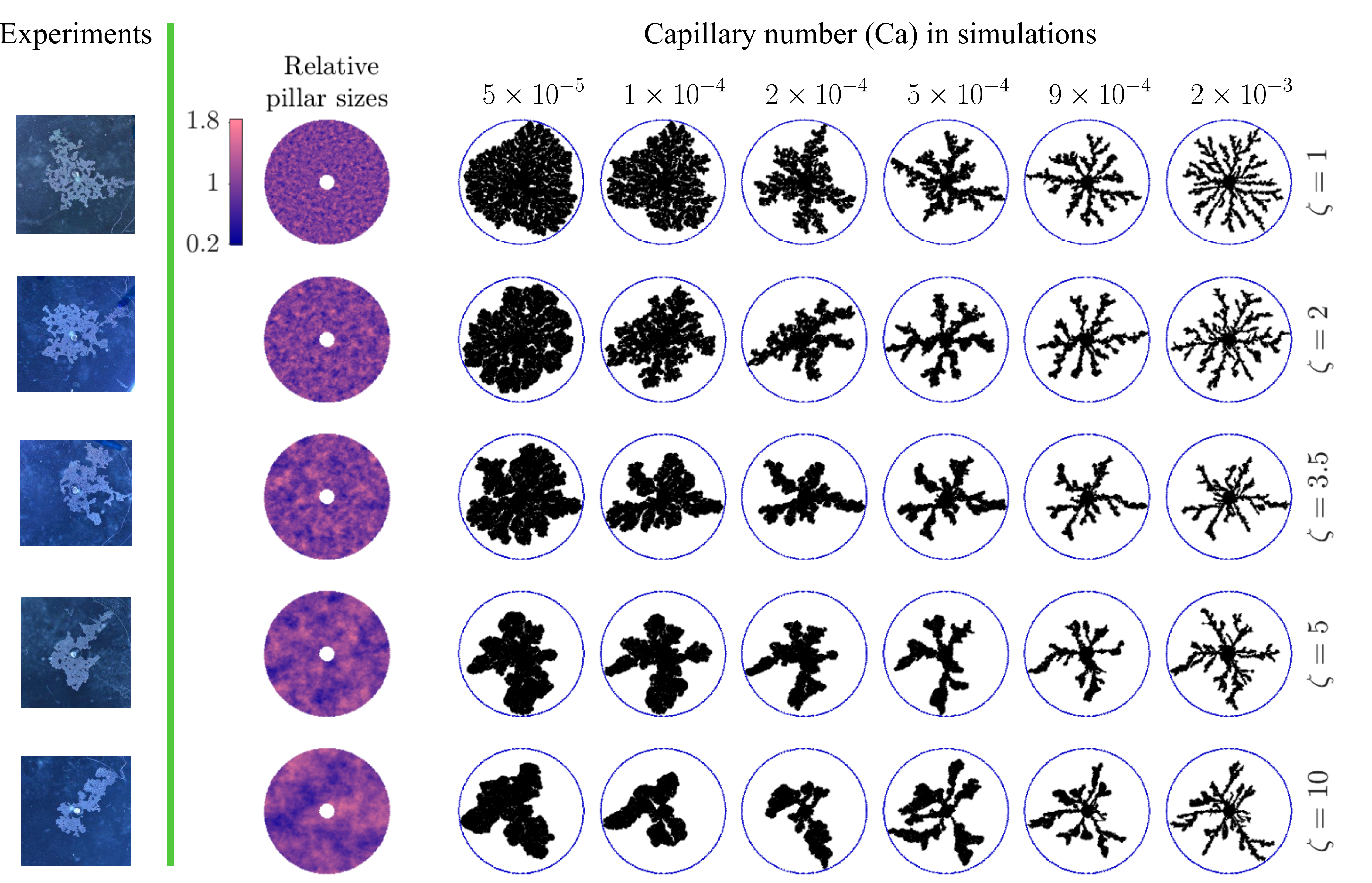} 
\caption{Displacement patterns at breakthrough in samples of various correlation lengths $\zeta$ and at different values of the capillary number $\mathrm{Ca}$.
Increasing $\zeta$ increases the size of regions or patches of similarly-sized pillars (e.g. a patch with larger openings and hence lower capillary thresholds and flow resistance), promoting preferential invasion through these regions, and resulting in patterns which follow more closely the underlying pore geometry.
The left-most column shows images of experiments at \(\mathrm{Ca}\)=\(1.3\times 10^{-5}\), where invading fluid (air) appears brighter.
The remaining results are from simulations, including representative maps of the relative pillar sizes for each $\zeta$.
Here, the invading and defending fluids appear in black and white, respectively, while the solid pillars are not shown.
The perimeter of the porous medium is denoted by a blue circle.
The same sample geometry, i.e. the set of \textit{relative} pillar sizes, is used for all displacement patterns shown in each row.
\label{fig:Results:SimPatterns}
}
\end{figure}

\subsection{Displacement patterns}
\label{sec:Results:Sim:Patterns}

Characteristic displacement patterns, both simulated and experimental, are presented in Fig.~\ref{fig:Results:SimPatterns}.
For the simulations, we show results at various flow rates (i.e. $\mathrm{Ca}$) for one specific realization (sample geometry) of each correlation length $\zeta$.
For slow flows ($\mathrm{Ca}\leq1\times 10^{-4}$), invasion is mostly controlled by the capillary invasion thresholds, determined directly from the sizes of the pore throats throughout the sample.
At intermediate rates ($\mathrm{Ca} = 2 \times 10^{-4}$ and $5 \times 10^{-4}$ ), fingers of the invading fluid become more apparent, especially for lower $\zeta$, as a result of the stronger viscous effects \citep{Meheust2002}.
Up to these rates, increasing $\zeta$ forces the displacement pattern to more closely follow the underlying pore geometry, invading mostly the patches of smallest pillars (hence, the largest pores).
Such similarities between the invasion patterns and maps of the pore sizes are evident in Fig.~\ref{fig:Results:SimPatterns}, for example.
Increasing $\zeta$ also leads to a small increase in the occurrence of cooperative pore invasion events, resulting in a smoother interface.

The impact of \(\zeta\) is relatively limited at high $\mathrm{Ca}$, where viscous fingering patterns inevitably emerge, as pressure screening inhibits invasion behind the most advanced edge of the displacement front~\citep{Lovoll2004}.
Nevertheless, increasing $\zeta$ reduces the number of invading fluid fingers, and lowers the displaced volume at breakthrough.  In many ways these invasion patterns resemble a skeleton or backbone of their low-rate analogues, showing that regions of large or small pores remain preferred locations for guiding or inhibiting the invading fingers, respectively.
Thus, their number is limited compared to the less correlated samples, in which fingers propagate equally freely in all directions.

The microfluidic experiments (Fig. \ref{fig:Results:SimPatterns}, left-most column) confirm our main findings from the simulations, at low Ca.
Here, increasing \(\zeta\) also leads to smoother and more preferential patterns, with reduced trapping of the defending liquid, and lower displaced volume at breakthrough.

\subsection{Interfacial features}
\label{sec:Results:Sim:Character}

\begin{figure}[ht!]
\centering
\noindent\includegraphics[width=1\columnwidth]{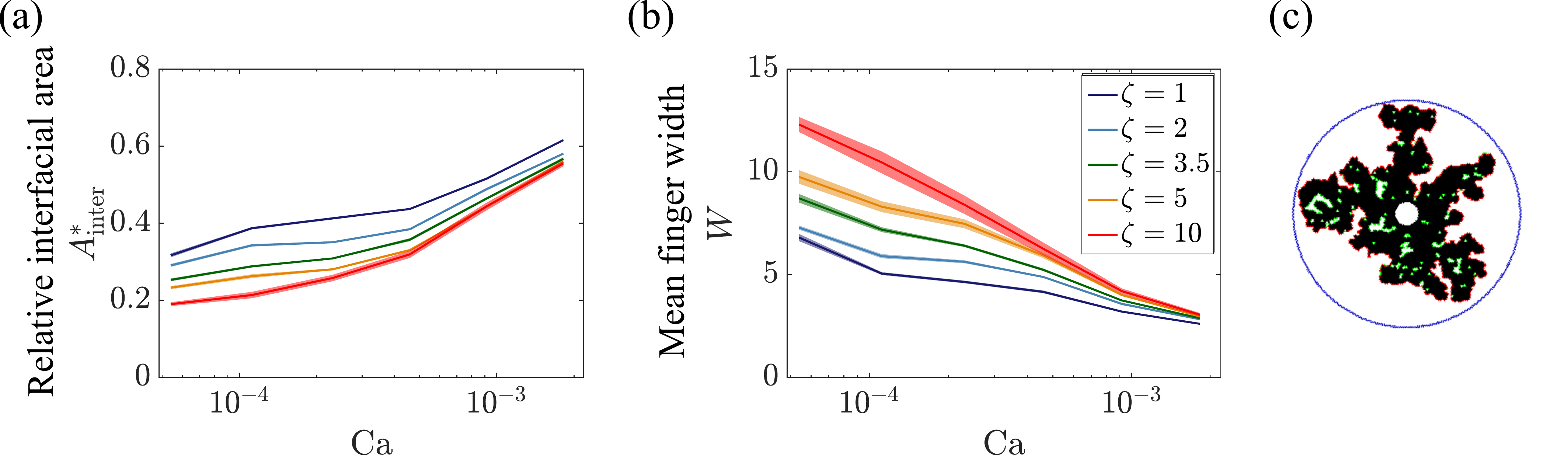} 
\caption{
Increasing the correlation length of the particle sizes \(\zeta\) reduces the relative interfacial area $A^*_\mathrm{inter}$ and increases the invading finger width \(W\).
(a) $A^*_\mathrm{inter}$ is lower for larger $\zeta$, as displacement patterns become smoother in response to longer-range correlations, and increases with the capillary number $\mathrm{Ca}$, as viscous forces become dominant.
(b) The finger width increases with \(\zeta\), and decreases with $\mathrm{Ca}$.
For each $\zeta$ we plot the ensemble average (lines) and the standard error (shading) of 10 independent realizations.
(c) We distinguish between the front area (marked in red)---the leading part of the interface only, excluding trapped clusters (green)---and the interfacial area, which is the sum of the front area and the perimeter of the trapped clusters.
Here the invading fluid is shown in black, and the cell perimeter in blue. 
For the effect of front area, see Supplementary Material.
\label{fig:Results:PatternChar}	
}
\end{figure}

Two prominent characteristics of fluid displacement patterns are the extensive interface between the invading and defending fluids, and the invading fluid's fingers \citep{Lovoll2004}. The interfacial area between the fluids is related to the pressure-saturation relationship \citep{Porter2010,Liu2011}, and can control the rates of fluid mixing and chemical reactions \citep{DeAnna2014}.
Here, we characterise it by $A^*_\mathrm{inter}$, which is the ratio of the total interfacial area (including trapped clusters) to the invaded volume, at breakthrough. 
As shown in Fig.~\ref{fig:Results:PatternChar}a, this relative area consistently decreases as the correlation length is increased, for all flow rates.
The largest effect is observed at low flow rates, reflecting a transition from capillary fingering at low $\zeta$ to smoother displacement patterns at high $\zeta$. In contrast, the increase in $A^*_\mathrm{inter}$ with $\mathrm{Ca}$ results from the transition towards viscosity-dominated patterns \citep{Lenormand1988,Holtzman2016}.

Another characteristic highlighting the transition from capillary to viscosity-controlled displacement is the increase in the relative \emph{front} area $A^*_\mathrm{front}$ with $\mathrm{Ca}$; here  $A^*_\mathrm{front}$ is defined as the ratio of the total front area (excluding trapped clusters, cf. Fig. \ref{fig:Results:PatternChar}c) to the invaded volume, at breakthrough.
This behavior occurs in a similar manner across all $\zeta$ values, as the increase in pore-size correlation brings the competing contributions of two phenomena---reduction in the invaded volume, and in the interface roughness (see Supplementary Material).

As shown in Fig. \ref{fig:Results:PatternChar}b, we find that the mean finger width, $W$, is generally higher for larger \(\zeta\).
This reflects the increasing size of the contiguous regions of large pores, through which the invasion proceeds at low $\mathrm{Ca}$, when capillarity is dominant.
At sufficiently high $\mathrm{Ca}$, viscous screening leads to the emergence of thin fingers, and minimizes the impact of the underlying porous microstructure (i.e. \(\zeta\)).
Here, $W$ converges to a value similar to that found in simulations of invasion into uncorrelated porous media \citep{Holtzman2016}, further exemplifying the reduced effect of microstructure, and specifically of correlations, at higher flow rates.
The width of an invading finger of fluid is measured here by averaging the widths of the shortest paths across the finger at each point along its interface (see skeleton-based algorithm in \citet{Holtzman2016}).

\begin{figure}[ht!]
\centering
\noindent\includegraphics[width=1\columnwidth]{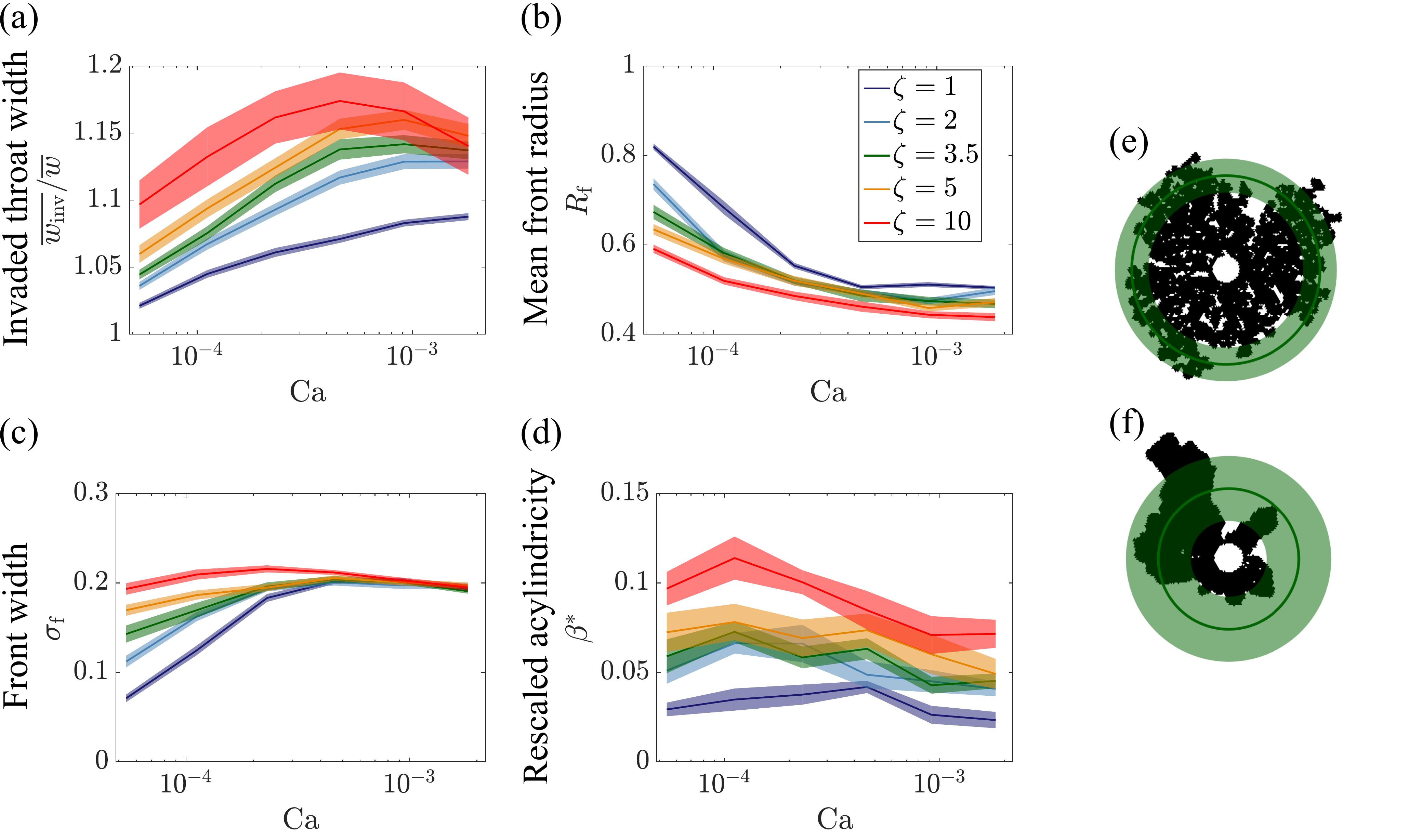} 
\caption{Spatial correlation leads to more selective, preferential and asymmetric displacement.
(a) At larger $\zeta$ the invasion is able to \textit{select} for a higher mean invaded throat width $\overline{w_\mathrm{inv}}/\overline{w}$. 
(b) This allows for a more \textit{preferential} route of the displacement front to breakthrough, and the mean radius of the displacement front, $R_\mathrm{f}$, decreases with both increasing correlation length $\zeta$ and capillary number $\mathrm{Ca}$.
(c) Similarly, the displacement front width, given by the standard deviation of the front position, $\sigma_\mathrm{f}$, increases with both $\zeta$ and $\mathrm{Ca}$. 
(d) Additionally, displacement patterns become less \textit{symmetric} with increasing $\zeta$, as quantified here by the rescaled acylindricity $\beta^*$, which is zero for a circular pattern.
For each $\zeta$ we plot the ensemble average (lines) and standard error (shading) of 10 realizations.
Example displacement patterns for $\zeta$=1 (e) and \(\zeta=10\) (f), at $\mathrm{Ca}=1\times 10^{-4}$, illustrate the definitions of the mean front radius $R_\mathrm{f}$ (dark green circle) and width $\sigma_\mathrm{f}$ (green shading).
For $\zeta$=1 the displacement front is further from the injection point on average (large $R_\mathrm{f}$), while the front positions are more narrowly spread out (small $\sigma_\mathrm{f}$), than for $\zeta$=10.
\label{fig:Results:PreferSelect}
}
\end{figure}

\subsection{Preferential displacement and invasion selectivity}
\label{sec:Results:Sim:PreferSelect}

We now turn to quantify the impact of correlations on how selective and preferential the invasion is.
We consider the invasion to be ``selective'' if it samples only a narrow range of the available pore sizes, as opposed to more randomly invading pores across the entire size distribution.
Similarly, we use the term ``preferential'' to describe a displacement advancing through distinct pathways or channels, rather than in a uniform radial front (i.e. certain routes are \textit{preferred}).
Selective invasion can result in preferential patterns, which is the case here for larger \(\zeta\):
by selectively invading through the connected regions of larger pores, the invading fluid propagates in more preferential patterns.

A selective invasion pattern will tend to favor the largest, easiest-to-invade pore throats.
As we show in Fig.~\ref{fig:Results:PreferSelect}, the mean width of the invaded throats (normalized by the mean throat width, i.e. $\overline{w_\mathrm{inv}}/\overline{w}$) increases with $\zeta$ for all flow rates, indicating that the correlations result in a more selective invasion.
The impact of the capillary number is less clear;
while the selectivity does also increase with $\mathrm{Ca}$, it saturates for faster flows, where viscous forces dominate, and may even decrease at the highest Ca studied.
As discussed in Section \ref{sec:Results:Sim:Patterns}, viscous fingers at high flow rates will still grow more readily into correlated regions of larger throats, and be inhibited by tighter pores.
Once any finger falls behind the main front, pressure screening will further limit its advance.
Thus, although finger propagation depends more on the viscous resistance to flow---a nonlocal feature, as opposed to the local capillary resistance---an interplay between these two resistance terms at the leading edge of the invasion front promotes selective invasion of the largest pores. 
Increasing $\zeta$ also promotes more preferential fluid displacement into the connected clusters of larger pores.  We characterize this with a series of metrics that describe whether the invasion front is smooth and symmetric (less preferential), or rough and asymmetric (more preferential).  For example, the average radius of the displacement front at breakthrough  decreases with increasing $\zeta$, as shown in Fig.~\ref{fig:Results:PreferSelect}b.
A high value of $R_\mathrm{f}$ (the mean front radius, scaled by the system size --- see Fig.~\ref{fig:Results:PreferSelect}e,f) indicates that the front has more evenly approached the perimeter in all directions.
Hence, a decrease in $R_\mathrm{f}$ shows that most of the displacement occurred through a smaller part of the cell, e.g. through fewer fingers. 
Another indication of the more preferential invasion at higher $\zeta$ is the increase in front width $\sigma_\mathrm{f}$---the standard deviation of the front location around $R_\mathrm{f}$ \citep{Prat1999}---as presented in  Fig.~\ref{fig:Results:PreferSelect}c.
The gradual transition from capillary fingering to viscous fingering, between intermediate and high $\mathrm{Ca}$, results in an increase of $\sigma_\mathrm{f}$ (i.e. a wider front, which is more pronounced for smaller $\zeta$) and a decrease of $R_\mathrm{f}$.

Another consequence of the more preferential and selective invasion with increasing $\zeta$, is that the patterns also become less symmetric.
This is quantified by the pattern acylindricity, $\beta^*$, which is calculated from the second moments of a best-fit ellipse to the invaded area \citep{Vymetal2011}.
Briefly, for a perfectly circular invasion pattern, $\beta^*=0$, whereas for a single needle-like growth along one direction, $\beta^*=1/3$ (for details of the calculation, see Appendix A).  
The highest values of $\beta^*$ are obtained for low-to-intermediate $\mathrm{Ca}$ and high \(\zeta\), when capillary forces dominate pore invasion, and the underlying heterogeneity has the greatest impact (Fig.~\ref{fig:Results:PreferSelect}d).
In brief, we have shown here that increasing $\zeta$ results in more selective fluid invasion, which leads to preferential displacement patterns;
these are characterized by lower front radius, higher front width, and lower symmetry.

\subsection{Displacement efficiency} \label{sec:Results:Sim:Efficency}

\begin{figure}[t!]
\centering
\noindent\includegraphics[width=0.85\columnwidth]{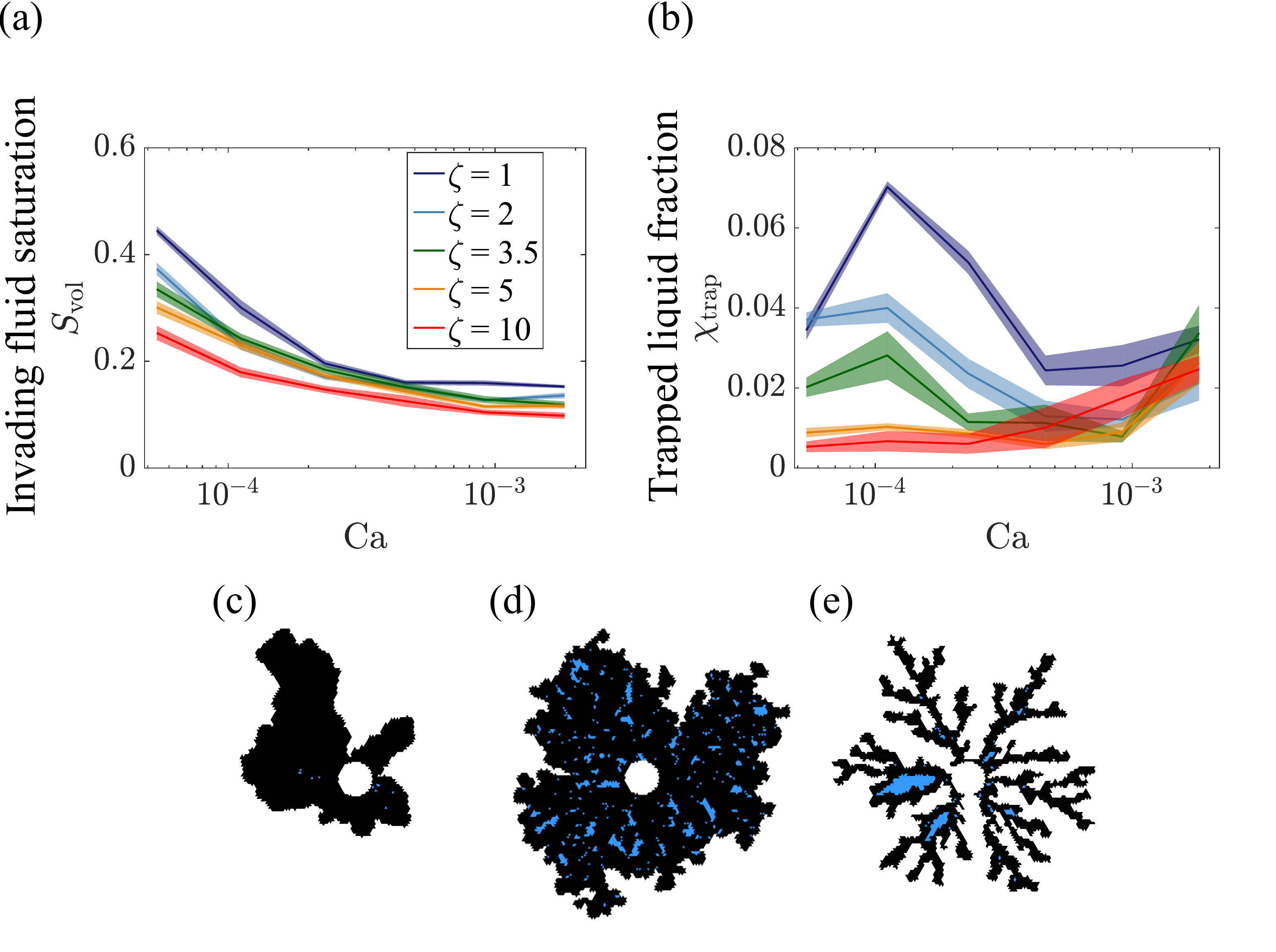} 
\caption{Effects of correlated heterogeneity on fluid displacement efficiency and trapping.
(a) The sweep efficiency---i.e. the breakthrough saturation $S$ of the invading phase---is reduced for longer-range correlations, due to the preferential invasion of larger pores. The effect of $\zeta$ is less apparent at high $\mathrm{Ca}$, where viscous fingering is responsible for low efficiency.
(b) The fraction of trapped defending liquid, $\chi_\mathrm{trap}$, is higher for shorter-range correlations, at most $\mathrm{Ca}$ values. 
Increasing $\zeta$  suppresses trapping under capillary-controlled invasion, as demonstrated by comparing displacement patterns for $\zeta$=10, $\mathrm{Ca}=1\times 10^{-4}$ (c) and $\zeta$=1, $\mathrm{Ca}=1\times 10^{-4}$ (d).
The increase in trapped liquid (blue regions) with $\mathrm{Ca}$ at viscous-dominated regime (high $\mathrm{Ca}$) is related to trapping \textit{between} viscous-controlled invading fingers (see panel e, for $\zeta$=1, $\mathrm{Ca}=2\times 10^{-3}$).
For each $\zeta$ we plot the ensemble average (lines) and standard error (shading) of 10 realizations.
\label{fig:Results:Efficiency}
}
\end{figure}

The sweep efficiency is one of the most important aspects of fluid displacement, from a practical standpoint, as it determines the ability to produce or withdraw fluids in applications such as oil production or groundwater remediation \citep{Wang2004,Ahmed2010}.
A typical measure of sweep efficiency is the invading fluid saturation at breakthrough, $S$.
Increasing $\zeta$ forces the displacement into fewer, more preferential pathways, and thus reduces $S$ (Fig.~\ref{fig:Results:Efficiency}a).
Due to the narrow, extended nature of viscous fingering, their emergence at high $\mathrm{Ca}$ also reduces significantly $S$, more sharply at lower $\zeta$.

The displacement efficiency is related to (and affected by) trapping---the isolation of defending fluid behind the displacement front in immobile, disconnected patches.
The trapped fraction $\chi_\mathrm{trap}$ depends in a complex manner on both the sample geometry and flow rate, leading to a non-monotonic dependence on $\mathrm{Ca}$ for most $\zeta$ values (Fig.~\ref{fig:Results:Efficiency}b). 
Here, $\chi_\mathrm{trap}$ is defined as the volumetric ratio of the trapped defending fluid to the total injected fluid, at breakthrough.
The non-trivial response of $\chi_\mathrm{trap}$ reflects a transition between three distinct trapping modes \citep{Holtzman2016}:
(i) For \textit{low} $\mathrm{Ca}$, the invading fluid efficiently fills the pore space, and trapping is limited (see case in Fig.~\ref{fig:Results:Efficiency}c);
(ii) at \textit{intermediate} $\mathrm{Ca}$,
capillary fingering patterns emerge, which trap multiple small islands of defending fluid (``capillary trapping'', Fig.~\ref{fig:Results:Efficiency}d);
(iii) at \textit{high} $\mathrm{Ca}$, viscous fingering becomes more dominant and trapping occurs in fewer, but larger, volumes in between distinct fingers (here called ``viscous trapping'', Fig.~\ref{fig:Results:Efficiency}e).
In the transition between capillary and viscous fingering (between intermediate and high \(\mathrm{Ca})\), trapping first decreases as capillary trapping becomes less efficient, and then rises again due to viscous trapping.
For the shorter-range correlations, the highest flow rates are characterized by more emerging invading-fluid fingers, compared to intermediate flow rates (see Fig \ref{fig:Results:SimPatterns}).
In these cases the higher number of fingers leads to more coalescence, and hence more trapping of the defending fluid, as compared to the simulations with long-range correlations.

We note that for the sample geometries and flow rates considered here, trapping is not the primary control on displacement efficiency.
This is mostly evident at high $\zeta$ and low $\mathrm{Ca}$, where the displacement occurs through few distinct regions;
while these regions are essentially contiguous, with a few small trapped islands, the preferential nature of the invasion pattern only allows it to explore a smaller section of the porous medium (compared to low $\zeta$) with a much lower overall efficiency.

\subsection{Comparing experiments and simulations} \label{sec:Results:Exp}

\begin{figure}[t!]
\centering
\noindent\includegraphics[width=0.85\columnwidth]{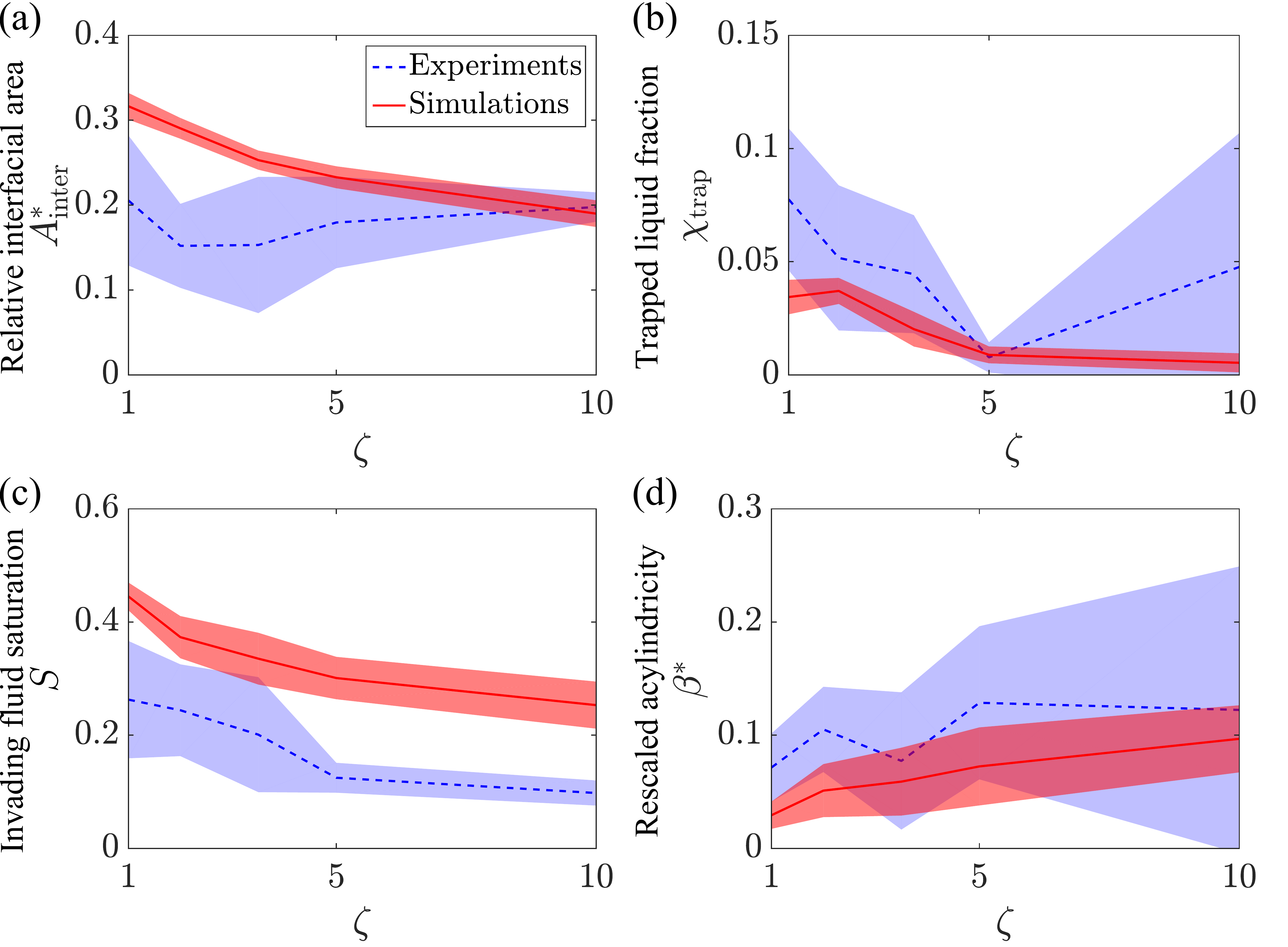} 
\caption{Our microfluidic experiments, conducted at lower $\mathrm{Ca}$, compare favorably with the simulations, in terms of trends as well as values of the following metrics:
(a) the interfacial area $A^*_\mathrm{inter}$,
(b) trapping fraction $\chi_\mathrm{trap}$,
(c) sweep efficiency $S$,
and
(d) rescaled acylindricity $\beta^*$.
We note that the simulations slightly overpredict $A^*_\mathrm{inter}$ (a) and $S$ (c).
Lines and shading represent averages and standard deviations, respectively, of 3--7 experimental realizations for each $\zeta$ (blue, at $\mathrm{Ca}$ between $ 2 \times 10^{-6}$ and $  4\times10^{-5}$) and 10 numerical realizations at each $\zeta$ (red, at $\mathrm{Ca} = 5 \times 10^{-5}$).
\label{fig:Results:ExpSimCompare}
}
\end{figure}

We used microfluidic fluid displacement experiments to validate our simulations.
As was shown in Fig.~\ref{fig:Results:SimPatterns}, the impact of the correlation length $\zeta$ on the resulting displacement patterns is similar in experiments and simulations, for the low $\mathrm{Ca}$ that are accessible experimentally.
Here, and in Fig.~\ref{fig:Results:ExpSimCompare}, we demonstrate how this agreement extends to a quantitative comparison of the following metrics:
the interfacial area, $A^*_\mathrm{inter}$,
trapped fraction, $\chi_\mathrm{trap}$,
sweep efficiency, $S$
and the dimensionless acylindricity, $\beta^*$.
These metrics are important for flow and transport, and can be reliably and consistently measured  from the experimental images.
As shown in Fig.~\ref{fig:Results:ExpSimCompare}, most metrics show similar magnitudes and trends in the experiments and simulations, with two minor exceptions.
First, while we find similar values of $A^*_\mathrm{inter}$ for experiments and simulations for $\zeta$ = 10, the decrease in $A^*_\mathrm{inter}$ with $\zeta$ is apparent in the simulations alone (Fig. \ref{fig:Results:ExpSimCompare}a).
The second small discrepancy is the consistently lower $S$ in the experiments vs. the simulations (Fig. \ref{fig:Results:ExpSimCompare}c).

In calculating the ensemble averages and standard deviation for each $\zeta$ in Fig. \ref{fig:Results:ExpSimCompare}, we average over all experiments in the range of the tested $\mathrm{Ca}$ values. The motivation behind this is to increase the number of experimental data values for each $\zeta$, improving the statistical significance of our analysis. This is justified by the small expected effect of $\mathrm{Ca}$ in the quasi-static limit (low $\mathrm{Ca}$) where viscous effects are negligible \citep{Toussaint2012}; results from all experiments presented in the online Supplementary Material confirm this assumption. 
While our current experimental setup limited us to low $\mathrm{Ca}$, we note that the impact of correlation is most significant at these $\mathrm{Ca}$ values, according to our simulations (Figs. \ref{fig:Results:PatternChar}--\ref{fig:Results:Efficiency}).  The close agreement in this limit thus adds particular strength to these results. 

\section{Discussion}
\label{sec:Discussion}

\subsection{Implications for viscous flow and solute transport} \label{sec:Discussion:TransportImplications}

The impact of spatial correlation on fluid displacement patterns, exposed in this study, implies that correlations can strongly influence additional aspects such as flow rates, solute transport and reaction rates.
The selective invasion of larger pores is characteristic of better correlated porous media (Fig.~\ref{fig:Results:PreferSelect}), and the resulting preferential displacement patterns increase the relative permeability of the invading phase (at a given saturation), as compared with uncorrelated media \citep{Rajaram1997}.
This effect is due to the control exerted on relative permeability by the fluids' spatial distribution and connectivity, changing, for instance, the constitutive relationship between relative permeability and saturation or capillary pressure \citep{Rajaram1997,Blunt2017}.
This is in line with previous observations of the effect of pore-size disorder (in uncorrelated samples) on fluid displacement patterns \citep{Holtzman2016} and its impact on relative permeability \citep{Li2018}.

Since solute transport is largely controlled by fluid flow, correlations in the microstructure can also lead to preferential solute transport pathways and localized reaction hotspots. A similar link between flow focusing and transport has been recently shown for uncorrelated heterogeneity \citep{Alim2017}, and it is to be expected that correlations would intensify this impact.
Specifically, our findings of the suppressive effect of pore-size correlations on the creation of fluid-fluid interfacial area (Fig.~\ref{fig:Results:PatternChar}) suggest a consequent effect on solute concentration gradients, solute mixing, and reaction rates \citep{DeAnna2014,Jimenez-Martinez2015} between the solutes carried by the invading fluid and those resident in the defending fluid.
For instance, in light of the positive effect of viscous fingering on fluid mixing and reaction rates \citep{Jha2011,LeBorgne2014,Nicolaides2015}, we expect that increasing correlations---by delaying the transition to viscous fingering---would reduce the mixing rate (at a given flow rate).
Transport heterogeneity, associated with differences in flow velocities between regions with high and low conductivity, was also shown to influence solute dispersion \citep{Kang2015,Pool2018}, again suggesting a link between pore-size correlations---and their effect on fluid transport---and solute transport.

Finally, the relationship between flow intermittency and pressure fluctuations during immiscible displacement was recently demonstrated in drainage experiments in uncorrelated porous media \citep{Moura2016a}, and in drying experiments and simulations in correlated media \citep{Biswas2018}.
The latter study also shows how increasing the correlation length leads to larger avalanches, as larger patches of similarly-sized pores become accessible per an increase in capillary pressure; this is consistent with our interpretation of increasing invasion selectivity.

\subsection{Environmental relevance of structural heterogeneity}
\label{sec:Discussion:EnvironmentalPhenom}

The effect of spatial correlations on fluid displacement, forcing it to become more
preferential, has implications for a variety of environmental processes at various length scales.
For example, it has been shown that structural heterogeneity can impact water infiltration in soils and the emergence of preferential flow \citep{Schluter2012}, the development of unsaturated zones at the river-aquifer interface \citep{Schilling2017}, and the distribution of saline and fresh water within the continental shelf \citep{Michael2016}.
The occurrence of structural heterogeneity, in the form of high permeability zones within aquitards, can also compromise their ability to act as barriers for water and contaminant migration \citep{Timms2018}.
In addition, the contrast in permeability between different geological layers may control the mechanisms for attenuation of CO$_2$, when considering leakage from an underground reservoir used for carbon capture and storage \citep{Neufeld2009,Plampin2017,Liang2018}.
These studies show the importance of considering the structural heterogeneity of porous materials, due to their impact on flow rates and fluid phase distribution. 
While upscaling pore-scale results to core or field scale remains a significant challenge \citep{Nooruddin2018}, pore-scale studies such as the one presented here serve as important building blocks for simulations of environmental phenomena at regional scales (e.g. \citep{Ebrahimi2018}).

\section{Summary and conclusions}

We have presented a systematic investigation of the impact of correlated heterogeneity on fluid displacement patterns, and its interplay with flow rates, in partially-wettable porous media, by combining pore-scale simulations with high-resolution microfluidic experiments.
We find that at low-to-intermediate flow rates (i.e. low or moderate $\mathrm{Ca}$ values, where capillarity dominates pore invasion) increasing the correlation length results in a lower sweep efficiency, reduced trapping of the defending fluid and lower interfacial area, with displacement patterns that are more preferential,
and follow more closely the underlying pore geometry. 
These patterns are further characterized by wider invading fingers and lower symmetry. 
At higher \(\mathrm{Ca}\), when viscosity dominates, we find that the impact of correlation becomes relatively limited, although the pattern symmetry and trapped fraction are still lower than for uncorrelated porous media.

Our results highlight the importance of the dynamic pore-scale modeling of multi-phase flow in porous media, which allows one to capture complex behavior that cannot be described by quasi-static models \citep{Paterson1996,Babadagli2000}, and which might otherwise completely evade coarse-grained, continuum  models \citep{Blunt2017}.
The spatial distribution of pore sizes and their connectivity affects the water distribution in the subsurface \citep{Schluter2012,Schilling2017}, transport, mixing and reaction of contaminants and nutrients \citep{DeAnna2014,Jimenez-Martinez2017,Pool2018}, and fluid displacement patterns in engineered porous materials \citep{Al-Housseiny2012,Rabbani2018}.  These are merely a few examples of applications in which structural heterogeneity plays a key role in fluid and solute transport across scales, emphasizing the need for models that properly incorporate pore-scale heterogeneity.

\section*{Acknowledgments}
This work was supported by the the State of Lower-Saxony, Germany (\#ZN-2823).
RH also acknowledges partial support from the Israeli Science
Foundation (\#ISF-867/13) and the Israel Ministry of Agriculture
and Rural Development (\#821-0137-13).
The authors thank Paolo Fantinel for assistance in designing the experimental system.

\appendix

\section*{Appendix A: Acylindricity and the gyration tensor}
\label{Sec:App:GyrTens}

The acylindricity \(\beta\) measures the divergence of a pattern from a circular shape, and is the difference between the eigenvalues \(T\) of the gyration tensor:
\begin{equation}
\beta = |T_1-T_2|.
\end{equation}
These eigenvalues give the squares of the lengths of the principal axes of the ellipse which best approximates the shape of the displacement pattern \citep{Vymetal2011}.  We thus use the radius of the cell, $L/2$, as a scale to give a dimensionless acylindricity of $\beta^*=4\beta/{L^2}$, which is 0 for a circular invasion pattern, and 1/3 for the extreme case of invasion proceeding along a straight line from the inlet to the outer perimeter.
The gyration tensor \(S\) is itself computed from the locations of the invaded pores, with respect to the inlet (the center of the cell),
\begin{equation}
S= 
\frac{1}{N}\sum_{i=1}^N
\begin{pmatrix}
x_{i}^2 & x_{i}y_{i}\\
y_{i}x_{i} &  y_{i}^2\\
\end{pmatrix},
\end{equation}
where $x$ and $y$ are the spatial coordinates, such that \(x=0\) and \(y=0\) at the inlet, and $i$ runs over all \(N\) invaded pores.

\bibliographystyle{elsarticle-harv} 
\bibliography{drain_corr_complete.bib}

\end{document}